\documentclass[%
 reprint,
superscriptaddress,
 amsmath,amssymb,
 aps,
]{revtex4-2}
\usepackage{graphicx}
\usepackage{dcolumn}
\usepackage{bm}
\usepackage{xcolor}

\begin{document}

\title[The effects on threshold induced correlations in RFIM]{The effects of external noise on threshold induced correlations in ferromagnetic systems}

\author{Dragutin Jovkovi\'{c}}
\affiliation{Faculty of Mining and Geology, University of Belgrade, POB 162, 11000 Belgrade, Serbia}

\author{Sanja Jani\' cevi\'{c}}
\affiliation{Faculty of Science, University of Kragujevac, POB 60, 34000 Kragujevac, Serbia} 

\author{Svetislav Mijatovi\'{c}}
\affiliation{Faculty of Physics, University of Belgrade, POB 44, 11001 Belgrade, Serbia} 

\author{Lasse Laurson}
\affiliation{Computational Physics Laboratory, Tampere University, POB 692, FI-33014 Tampere, Finland} 

\author{Djordje Spasojevi\'{c}}
\affiliation{Faculty of Physics, University of Belgrade, POB 44, 11001 Belgrade, Serbia}

\begin{abstract}
In the present paper we investigate the impact of the external noise and \textcolor{black}{detection} 
threshold level on the simulation data for the systems that evolve through metastable states. 
As a representative model of such systems we chose the nonequilibrium athermal random field 
Ising model with two types of the external noise,
\textcolor{black}{
uniform white noise and Gaussian white noise with various different standard deviations, 
imposed on the original  response signal obtained in model simulations.
} 
We applied a wide range of detection threshold
 levels in analysis of the signal and show how these quantities affect the values of exponent $\gamma_{S/T}$ 
 \textcolor{black}{(describing the scaling of the average avalanche size with duration),} the shift of waiting 
 time between the avalanches, and finally the collapses of the waiting time distributions. The results are
  obtained via extensive numerical simulations on the equilateral three-dimensional cubic lattices of various sizes and disorders.
\end{abstract}

\maketitle

\section{Introduction}\label{intro}
Although modern equipment is capable of significantly reducing the superposition of unwanted external noise on experimental data, still this impact cannot be avoided. Obviously, in order to optimally capture the observed phenomena one would like to lower as much as possible the external noise and the noise from the measurement system. Under the foregoing conditions the studied events are usually recognized as the parts of recorded signal lying above/below the  upper/lower threshold level imposed on the base line level (i.e. an idealized signal level in the absence of response signal and all types of noise). In this way, the events, thresholds, and noise are intertwined in each experimental signal.

Measurements on the systems exhibiting avalanche-like relaxation are not an exception. A vast diversity of phenomena dominantly evolving via avalanche-like events can be found in everyday life, e.g. earthquakes \cite{Earthquakes}, neuronal activities \cite{NeuronalAvalanches2012,BrainSignalsPRL2006}, financial markets \cite{Financial2013}, crystalline \cite{Zaiser,Lase2020,Lase2018,Lase2014,SanjaJStat} and amorphous \citep{Budrikis,Sandfeld} solids undergoing plastic deformation, cracks propagating in disordered solids \citep{Santucci,SanjaPRL}, etc. The mentioned type of relaxation  can cause the extreme events such as avalanches that span almost the whole system leading to a phase transition in the thermodynamic limit \cite{BelangerNatterman}. Among these phenomena, magnetization and relaxation processes in ferromagnetic materials play a distinguished role \cite{Lieneweg1972,Stanley1996,Durin2000,Kim2003,Shin2007,Ryu2007,Benassi-Zapperi-2011,Lima2017,Bohn2018}.  

In order to model and explain the Barkhausen noise that emerges when the ferromagnetic sample is driven by varying external magnetic field, a number of theoretical models were developed \cite{Cizeau1998,Zapperi1998,Sethna2006,SethnaJMMM2001,Bertotti1990,Franz2011,Vives1994,VivesJMMM2000,ABBM,DW,LassePRB2014}. One of the most prominent appears to be the random field Ising model (RFIM), that has been extensively studied in the past few decades \cite{Schulz1988,TadicPRL1996,BalogPRB2018,SethnaPRL93,Vives2001,FytasPRL2013,JSTAT2021}. Renormalization group approach has brought
certain answers regarding the RFIM critical behavior,
but it turned out to be a rather difficult task. The results obtained via perturbative renormalization group showed the limits of this approach as some incorrect predictions in three dimensions arose \cite{YoungPerturbation,ParisiPerturbation,BricmontPerturbation}. This led to the non-perturbative approach that appeared to give better results \cite{ParisiNonPerturbation,TissierNonPerturbation}. Recent numerical investigation of equilibrium version of RFIM offered important information on behavior of this type of model \textcolor{black}{\cite{FytasPRL2016,FytasPRE2017,FytasPRL2019}}.

The nonequilibrium version of RFIM turned out to be more relevant for the correspondence with experiments due to its locally driven dynamics. The nonequilibrium model provides temporal evolution mimicking the response of real ferromagnetic samples to the varying external magnetic field. This model has been studied numerically in a lot of papers. Its critical behavior and scaling laws in the case of equilateral lattices were investigated in \cite{OlgaPRB1999,DahmenPRB1999,RechePRB2003,RechePRB2004,SpasojevicEPL2006,SpasojevicPRL2011,SpasojevicPRE2011,SpasojevicPRE2014, JSTAT2021}, whilst recently the systems with different geometry were studied in \cite{NavasVives,3D-2D2018,BosaSciRep2019,3D-2DHceff,SvetaPRE2020} together with the impact of lattice topology on its criticality \cite{Shukla2013PRE,Shukla2015PRE,Shukla2016,Triangle,Shukla2019}.

So far, very few studies were done on the joint effect that both threshold and noise have on signals obtained from ferromagnetic materials. Recent experimental \cite{BohnExp} and theoretical \cite{SanjaSciRep} studies delivered some important results caused by the implementation of the finite detection threshold when analyzing the original signal.
\textcolor{black}{Effects of thresholding have been considered also, e.g., in the context of fracture \cite{bares2019seismiclike,SanjaPRL}, and argued to be of importance in seismicity \cite{post2021interevent,radiguet2016triggering}. Moreover, the problem has been studied also, e.g., in the case of birth-death processes \cite{font2015perils}. However, the joint effects due to thresholding a crackling noise signal with superimposed additive external white noise remain largely unexplored.}

In the present paper we investigate the joint impact of the external noise and imposed threshold level on the avalanche statistics extracted from the simulations of the nonequilibrium athermal RFIM on equilateral three-dimensional cubic latices of size $L$ containing $L^3$ spins. In this work we used two types of noise. One is \textcolor{black}{uniform} white noise 
taken from the uniform distribution (UWN) of width $w$, i.e. the noise that takes with probability density $p(n)=1/2w$ any value $n$ from the interval $[-w,w]$ and has the standard deviation $\sigma=w/\sqrt{3}$. 
The other type is the white noise taken from the zero mean Gaussian distribution with standard deviation $\sigma$ (GWN). 
\textcolor{black}{
These two theoretically convenient types of noise with flat  power spectral density $S(f)=const$ are (almost) 
ubiquitous in experiments (e.g. UWN as the quantization noise and GWN as electronic noise arising in amplifiers and detectors), 
and in many instances superposed by some $1/f$ noise, having  power spectral density $S(f)\propto 1/f^{\alpha_n}$ and 
various origins (see e.g. \cite{Weissman}), whose influence we defer for later studies.
}

The paper is organized as follows. Description of the model, together with the simulation details and algorithm description, are given in Section \ref{model}. 
In Section \ref{threshold} is explained what is achieved by thresholding of the signal, while in the Section \ref{impact} is shown how addition of external noise affects the properties of the relevant statistics. Finally, in Section \ref{conclusion} we give a discussion and conclusion of this study.

\section{Model}\label{model}

The RFIM is defined as follows. At each site $i$ of the underlying lattice lies the spin $S_i$ having two possible values $\pm 1$. There are three types of interaction to which the spins are exposed: 1) they interact with the nearest neighbours via exchange interaction, 2) there is the interaction between each spin and the applied external magnetic field $H$, and 3) every spin $S_i$ interacts with a local random field $h_i$ at its site. These random field values are chosen independently and without site-to-site correlations from some zero mean distribution so that the average taken over all possible random field configurations satisfies $\langle h_i h_j \rangle = R^2\delta_{i,j}$, where $\delta_{ij}$ is the Kronecker delta function, and $R$ is disorder, i.e. the standard deviation of the employed random field distribution. One such distribution is the Gaussian distribution $$\rho (h)=\frac{1}{\sqrt{2\pi}R}exp\Big(-\frac{h^2}{2R^2}\Big)$$ used in this paper.
Taking all three interactions into account the Hamiltonian of the system reads
\begin{equation}
    \mathcal{H}=-J\sum_{\{i,j\}} S_iS_j-H\sum_i {S_i}-\sum_i {h_iS_i}.
    \label{Hamiltonian}
\end{equation}
In the first term $J$ represents the strength (1 in this paper) of ferromagnetic coupling between the nearest neighbors, and the summation is performed over all distinct pairs $\{i,j\}$ of such spins.
The system behavior is governed by the local relaxation rule meaning that the spin $S_i$ is stable while its sign is the same as the sign of the \textit{effective field}$$h_i^\mathrm{eff}=\sum_{\langle j\rangle}S_j+H+h_i$$
where the summation in the first term is performed over all nearest neighbors of the spin $S_i$.
All spins that are unstable at the current moment will flip in the next moment of discrete time affecting neighbouring spins in a way that they can become unstable and flip in the next-next moment. This explains the mechanism for creation of avalanches. During the started avalanche the external field is kept constant and afterwards increased in a single step exactly to the value that will flip the least stable spin. This regime is known as \textit{adiabatic}. Each simulation begins with $H=-\infty$ and all spins being $-1$, and stops when all spins have value $+1$. All simulations are done with periodic boundary conditions along all three directions.

As already mentioned, while an avalanche is active we check all the nearest neighbors of the spins flipped at the moment $t$ and those of them that are unstable we flip in the moment $t+1$. In the simulations this a very fast process  whereas the finding of the next spin to be flipped once the avalanche is over is the most time-consuming. The so-called brute force algorithm \cite{Kuntz1999} suggests to check all non flipped spins in the system and find which one is the least stable. In big systems, like are ours in this paper, the time needed for such search is extremely large. In order to decrease the time consumption we use, therefore, the sorted list algorithm \cite{OlgaCondMat96,Kuntz1999}, which we implemented in Fortran.

\textcolor{black}{
Our results are obtained in extensive simulations of system sizes up to $L=1024$ for disorders $R$ surpassing the effective critical disorder $R_c^{\mathrm{eff}}(L)$ 
which provides only the nonspanning avalanches which are mostly encountered in experiments, see \cite{JSTAT2021}.
}
The results gathered from the simulations were analyzed using the proprietary programs coded in Fortran, Visual Basic and Wolfram Mathematica.

\section{Thresholding of the signal}\label{threshold}
\textcolor{black}{The nonequilibrium RFIM systems belongs to the class of systems whose response signal $V(t)$ when driven by the increasing external magnetic field is equal to the number of spins flipped at the moment $t$. In what follows we will limit our analysis to the systems having the response signal $V(t)>0$ while the system is active and $V(t)=0$ otherwise.}  
Unlike the signals generated in simulations, where the overall registered signal is  $V(t)$, the overall signals registered in experiments contain external noise $n(t)$. In this case, the registered signal is $V(t)+n(t)$, so it  becomes much harder to extract events from the registered signal corresponding to individual avalanches. 
One of the extraction methods is to impose some threshold level $V_{\mathrm{th}}$ and observe only the activity above $V_{\mathrm{th}}$, dividing the signal into sub-avalanches that are parts of some underlying avalanche. During an avalanche a sub-avalanche starts at the first moment $t_s$ when the signal surpasses the chosen threshold level, $V(t_s)+n(t_s)>V_{\mathrm{th}}$, and ends at the first moment of time when the signal falls below it, i.e. at the moment $t_e$ when $V(t_e)+n(t_e)\le V_{\mathrm{th}}$. The difference between these two moments is the duration of that sub-avalanche, $T=t_\mathrm{e}-t_\mathrm{s}$, while the size of the sub-avalanche is defined as the area that lies between the signal and the imposed threshold, $S=\int_{t_\mathrm{s}}^{t_\mathrm{e}}dt[V(t)-V_{\mathrm{th}}]$.

\begin{figure}[htpb]
    \centering
    \includegraphics[width=0.5\textwidth]{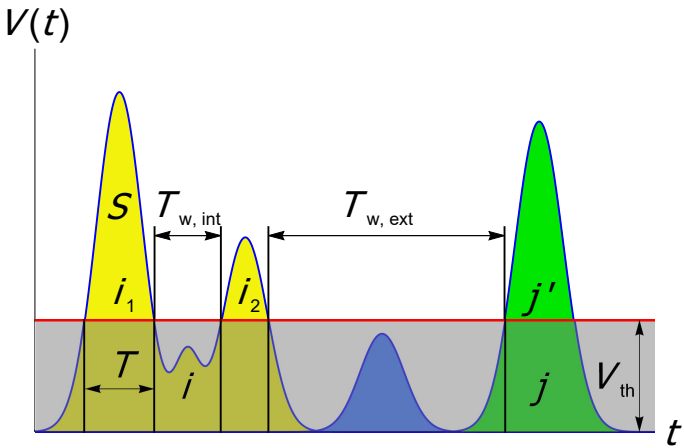}
    \caption{\textcolor{black}{An example illustrating the imposing of threshold $V_{\mathrm{th}}$ on the response signal, extraction of 
sub-avalanches, and definition of internal and external waiting time, $T_{\mathrm{w, int}}(V_{\mathrm{th}})$ and $T_{\mathrm{w, ext}}(V_{\mathrm{th}})$.
Two avalanches (labeled by $i$ and $j$) surpassing the threshold are separated by one (blue-colored) avalanche lying below the threshold. Two 
sub-avalanches $i_1$ and $i_2$ are extracted from the avalanche $i$, and a single sub-avalanche $j'$ from the avalanche $j$. Everything lying below $V_{\mathrm{th}}$ including parts of the avalanches $i$ and $j$ 
lying below the threshold is greyed.}   
}
    \label{Fig1}
\end{figure}

The thresholding process defined in this way introduces concept of waiting time as the time between two consecutive sub-avalanches.
\textcolor{black}{ 
In simulations one can recognize two basic kinds of waiting time illustrated in  Fig. \ref{Fig1}: 
the internal waiting time  $T_{\mathrm{w, int}}(V_{\mathrm{th}})$ between two consecutive 
sub-avalanches (yellow-colored and labeled by $i_1$ and $i_2$) that belong to the same ongoing avalanche 
(labeled by $i$)  
and the external waiting time $T_{\mathrm{w, ext}}(V_{\mathrm{th}})$ elapsed between the 
end of the last sub-avalanche $i_2$ from the ongoing avalanche $i$ and the start of the next (green-colored) sub-avalanche $j'$ 
belonging to the first succeeding avalanche $j$ surpassing the threshold, see \cite{AdiabaticReg}.
} 
 
Since it is known that external noise can have considerable impact on the properties defined by the thresholding process, our goal was to investigate that impact. To this end we thresholded generated simulation signals with superimposed noise. 
 
\section{The effect of adding the noise}\label{impact}
In this section we present the impact of adding two types of white noise, uniform and Gaussian, 
on the distributions of average avalanche size and properties of the internal and external waiting times.
\textcolor{black}{
The added noise is of external origin (e.g. noise that in experiments originates from detectors, amplifiers, AD converters, ambient EM interference, etc.) 
and is considered here to be much more pronounced than the system's intrinsic (e.g. thermal) noise. This, in particular, means that 
the system intrinsic dynamics is (practically) not disturbed by such noise and that the noise solely affects the registered signal by superposing on 
the pristine signal, i.e. the signal that would be registered by an ideal experimental system.
}

\subsection{Average avalanche size}\label{gamma}
\begin{figure}[htpb]
    \centering
    \includegraphics[width=0.5\textwidth]{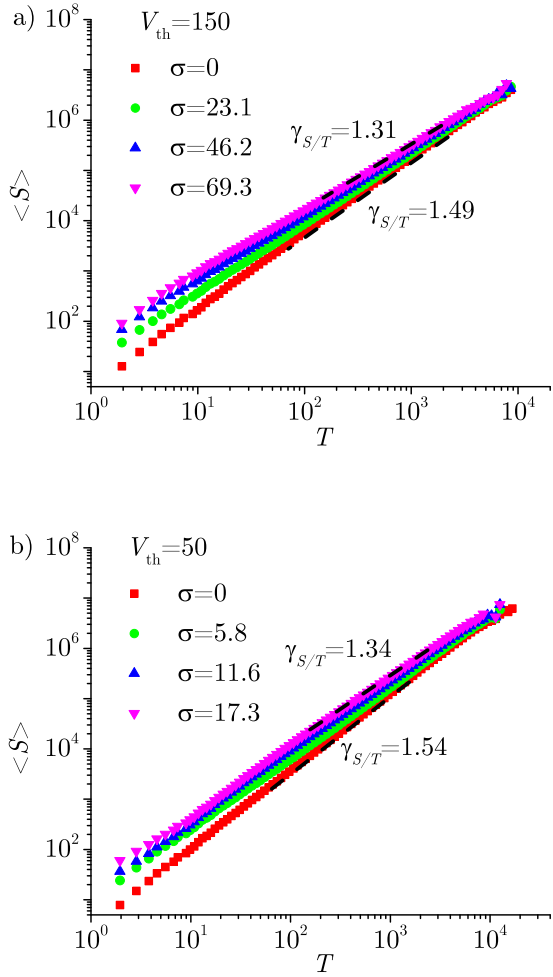}
    \caption{Average avalanche size for a given avalanche duration for the system of size $1024\times 1024\times 1024$ and disorder $R=2.25$. Detecting threshold level is $V_{\mathrm{th}}=150$ in UWN case when noise function standard deviations ranges from 0 to $69.3$ (panel a), and $V_{\mathrm{th}}=50$ in GWN case when noise function standard deviation ranges from 0 to $17.3$ (panel b).}
    \label{Fig2}
\end{figure}

\begin{figure}[htpb]
    \centering
    \includegraphics[width=0.5\textwidth]{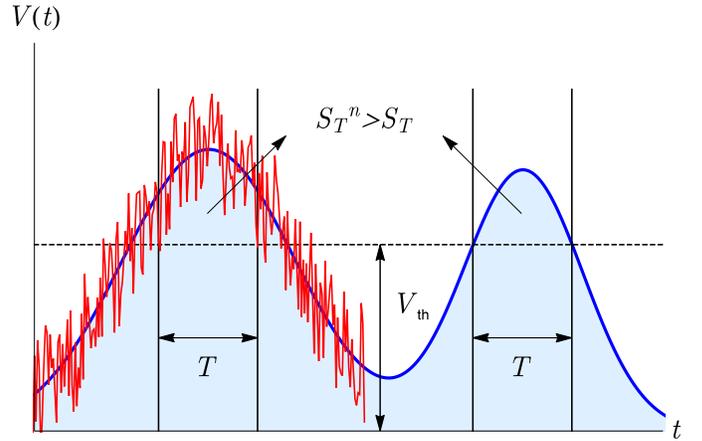}
    \caption{Example of the noise effect on the $\gamma_{S/T}$ values decrease. On the left side of the figure is presented the part of the signal with added noise (red line), while on the right side is the signal without noise (blue line). At a given threshold level for the same duration we see that the area (i.e. avalanche size) $S_T^n$, when the noise is applied, is larger than the area $S_T$ without the noise.}
    \label{Fig3}
\end{figure}

In Fig. \ref{Fig2} are present against duration $T$ the average size $\langle S\rangle_T$ of avalanches having duration $T$. This is done for added a) UWN and b) GWN to the original signal with the threshold levels $V_{\mathrm{th}}=150$ and $V_{\mathrm{th}}=50$, respectively. Average avalanche size data follow the power law $\langle S\rangle_T\sim T^{\gamma_{S/T}}$ 
\textcolor{black}{
specified by the universal RFIM exponent $\gamma_{S/T}$ appearing also as the exponent of the power spectral density of the RFIM signals, 
$S(f)\propto 1/f^{\gamma_{S/T}}$ \cite{Kuntz2000}.
}
We see that for both types of noise the slope of the $\langle S\rangle_T$ curves decreases as the noise standard deviation grows. This decrease  happens because the noise cuts long avalanches into shorter subavalanches. Since those shorter sub-avalanches originate from a longer one having larger average signal $\langle V(t)\rangle=\langle S \rangle /T$, their sizes are likely to be larger than the sizes of the regular avalanches of the same duration. In Fig. \ref{Fig3} we illustrate an example of how the avalanche of the same duration has greater size when the external noise is present in the signal. As the avalanche duration grows the previously explained effect becomes less expressed. Thus, the slope on the log-log plot of $\langle S\rangle$ versus $T$ curve, and therefore the $\gamma_{S/T}$ values, decreases with the increase of the noise standard deviation. 

\begin{figure}[htpb]
    \centering
    \includegraphics[width=0.5\textwidth]{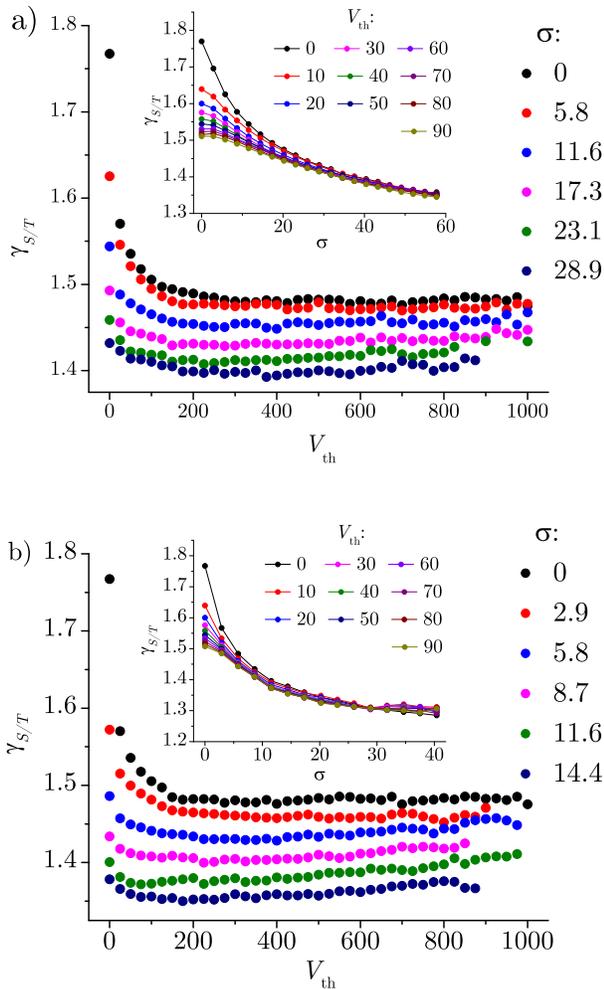}
    \caption{a) Main panel: values of $\gamma_{S/T}$ versus detecting threshold level for the standard deviation of UWN from $\sigma=0$ to $\sigma=28.9$. Inset: Values of $\gamma_{S/T}$ for wider range of $\sigma$ and for $V_{\mathrm{th}}$ that ranges from 0 to 90 (i.e. the values of $V_{\mathrm{th}}$ before the plateau). b) Main panel: values of $\gamma_{S/T}$ versus detecting threshold level for the standard deviation of GWN from $\sigma=0$ to $\sigma=14.4$. Inset: The same as in inset of panel a).}
    \label{Fig4}
\end{figure}

We present how $\gamma_{S/T}$ behaves for various threshold values and noise standard deviations $\sigma$ for UWN in Fig. \ref{Fig4}a) and GWN in Fig. \ref{Fig4}b). In the main panels of both Fig. \ref{Fig4}a) and \ref{Fig4}b) we see that the values of $\gamma_{S/T}$ drop quickly when we increase the threshold from zero, whilst after some value of $V_{\mathrm{th}}$, there is a plateau, i.e. a wide range of $V_{\mathrm{th}}$ for which the values of exponent $\gamma_{S/T}$ remain constant. The plateau is present because at these $V_{\mathrm{th}}$ values the impact of the originally small avalanches cannot be seen \cite{SanjaSciRep}. However, if $\sigma$ is large enough the difference in $\gamma_{S/T}$ disappears even for the small threshold values. The reason for this lies in the fact that regardless the value of $V_{\mathrm{th}}$, the average size of avalanches of small duration is dominantly governed by the noise. The average size of long avalanches in any case is not much affected by threshold or noise. So we expect that for the large noise standard deviation, $\gamma_{S/T}$ remains the same independently of the threshold. This can be observed in the insets of Fig. \ref{Fig4}, where we show how the values of $\gamma_{S/T}$ change for $0\leq \sigma \leq 60$ and $0\leq V_{\mathrm{th}}\leq 90$. We notice that with increase of $\sigma$ the deviations between the $\gamma_{S/T}(V_{\mathrm{th}})$ curves corresponding to different thresholds vanish. 

\begin{figure}[htpb]
    \centering
    \includegraphics[width=0.5\textwidth]{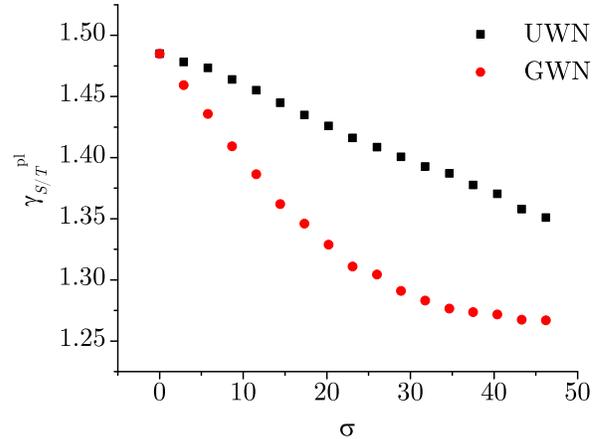}
    \caption{Plateau values of $\gamma_{S/T}$ obtained from Fig. 4 for various noise standard deviations. Black squares represent the results for UWN, while the red circles are the results for GWN.}
    \label{Fig5}
\end{figure}

In Fig. \ref{Fig5} we present the values of $\gamma_{S/T}$ at plateaus, denoted by $\gamma_{S/T}^\mathrm{pl}$, for various noise standard deviations for both UWN and GWN. It seems like that there is a wide range of linear decrease of the $\gamma_{S/T}^\mathrm{pl}$ with the $\sigma$ increase in the UWN case, whereas for the GWN case that range is smaller, after which the values of $\gamma_{S/T}^{\mathrm{pl}}$ saturate. Still, we have no explanation for such behavior.

\subsection{Waiting times}\label{shift}
\begin{figure}
    \centering
    \includegraphics[width=0.5\textwidth]{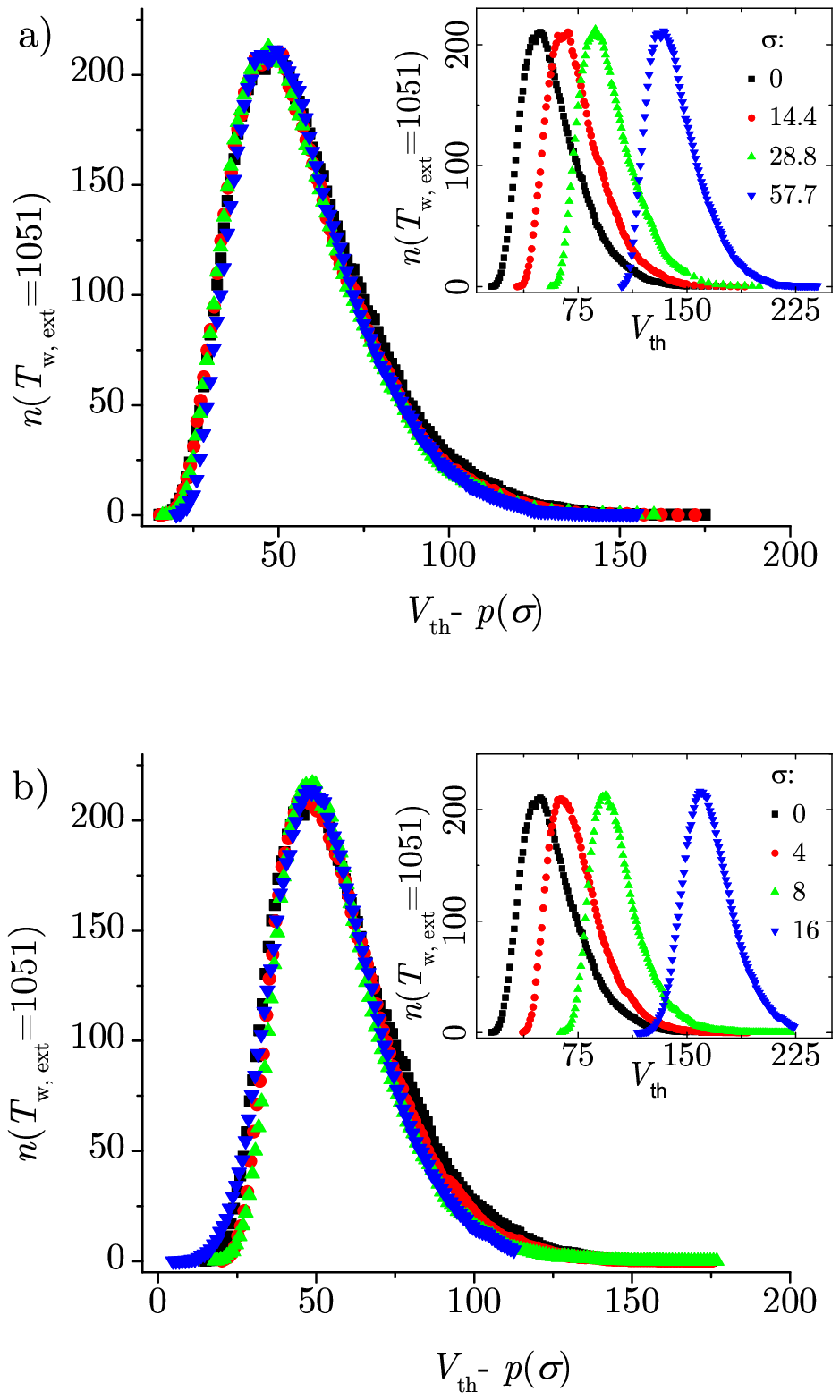}
   \caption{Number of occurrence of the external waiting time value $T_{\mathrm{w, ext}}=1051$ versus detection threshold level $V_{\mathrm{th}}$ for various standard deviations varying from $\sigma=0$ to $\sigma=57.7$. As shown in the main parts of both panels, the distributions collapse onto a single curve when presented against the threshold displaced by the shift parameter $p(\sigma)$ depending on the noise standard deviation $\sigma$. The data is obtained for the $1024\times1024\times1024$ system at $R=2.40$.}
    \label{Fig6}
\end{figure}

\begin{figure}
	\centering
	\includegraphics[width=0.5\textwidth]{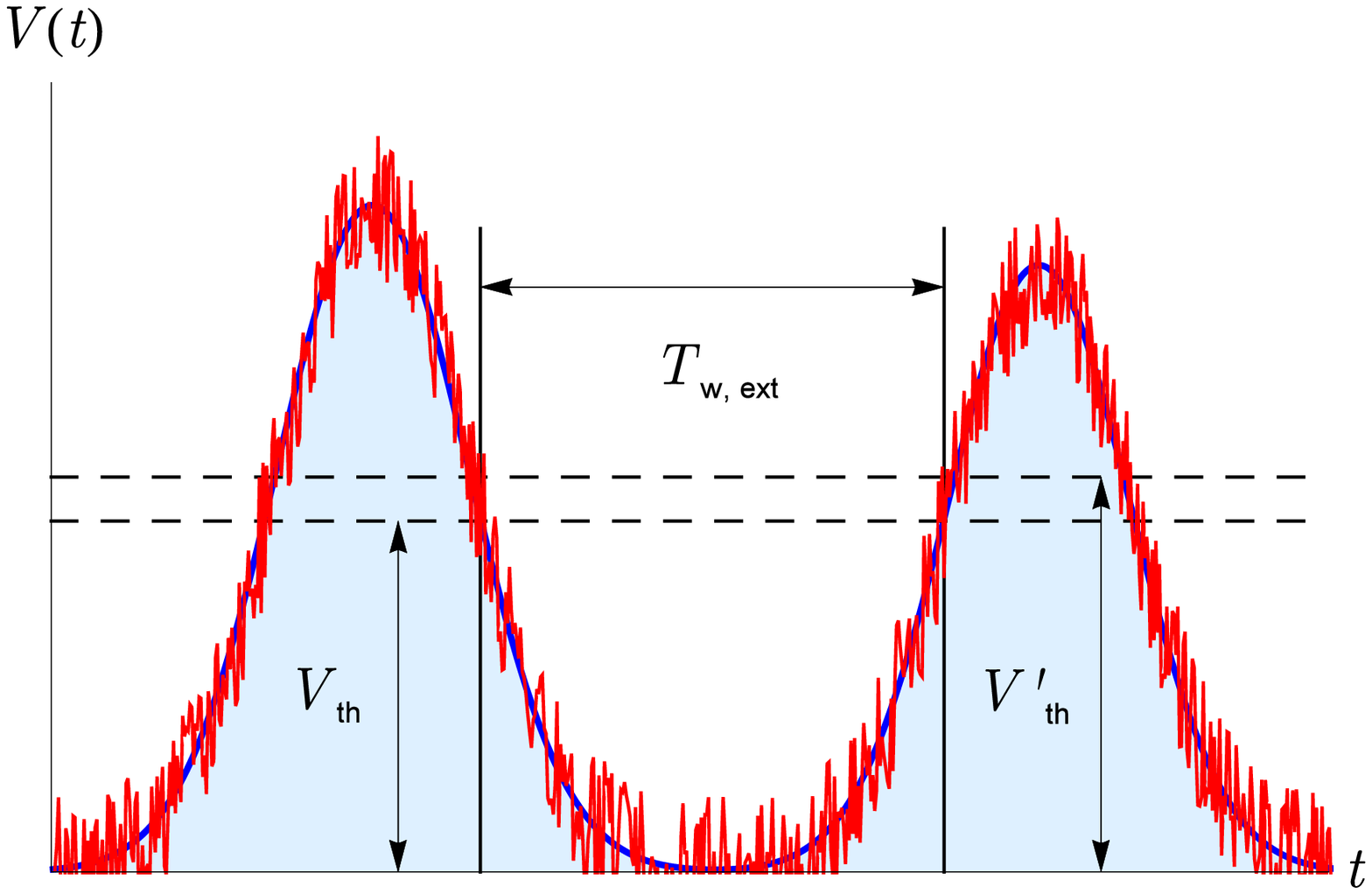}
	\caption{Schematic illustration of the average effect of added noise on the external waiting time: the same value of the external waiting time $T_{\mathrm{w, ext}}$ is found at a higher threshold level $V'_{\rm th}$ in the response signal with added noise (red line) compared to the threshold level $V_{\mathrm{th}}$ corresponding to the same waiting time for the pure response signal (blue line).}
	\label{Fig7}
\end{figure}

We start by observing the number of occurrences \textcolor{black}{$n(T_{\mathrm{w}}; R, V_{\mathrm{th}}, \sigma,L)$} \cite{LasseJStat} 
of waiting time $T_{\mathrm{w}}$ in one run at the threshold level $V_{\mathrm{th}}$ imposed on the response signal $V(t)$ with 
added noise of standard deviation $\sigma$. 
In insets of Fig. \ref{Fig6}, \textcolor{black}{showing a representative example  obtained 
for $T_{\rm w}=1051$, $L=1024$, and $R=2.40$,} one can see that the graphs of these distributions at fixed values of 
$\sigma$ shift to the right as $\sigma$ grows. 
This happens because on average the added noise increases the threshold 
level (at which a certain waiting time is found in a given section of recorded signal) from the value $V_{\mathrm{th}}$ for 
the pure signal to $V_{\mathrm{th}}'$ for the signal with noise, as is illustrated in Fig. \ref{Fig7}. In the case of 
\textcolor{black}{$n(T_{\mathrm{w}}; R, V_{\mathrm{th}},\sigma,L)$} distribution, we found that the mentioned increase 
can be described by a shift parameter $p(\sigma)$ depending on the standard deviation $\sigma$ of the applied noise. 
More specifically, if distribution \textcolor{black}{$n(T_{\mathrm{w}}; R, V_{\mathrm{th}}', \sigma,L)$} is translated 
along the threshold axis by the amount $p(\sigma)$, it overlaps with the distribution 
\textcolor{black}{$n(T_{\mathrm{w}}; R, V_{\mathrm{th}}, \sigma=0,L)$}:
\textcolor{black}{
\begin{equation}
n(T_{\mathrm{w}}; R, V_{\mathrm{th}}'-p(\sigma), \sigma, L)=n(T_{\mathrm{w}}; R, V_{\mathrm{th}}, \sigma=0, L).
\end{equation}
}

\begin{figure}
    \centering
    \includegraphics[width=0.5\textwidth]{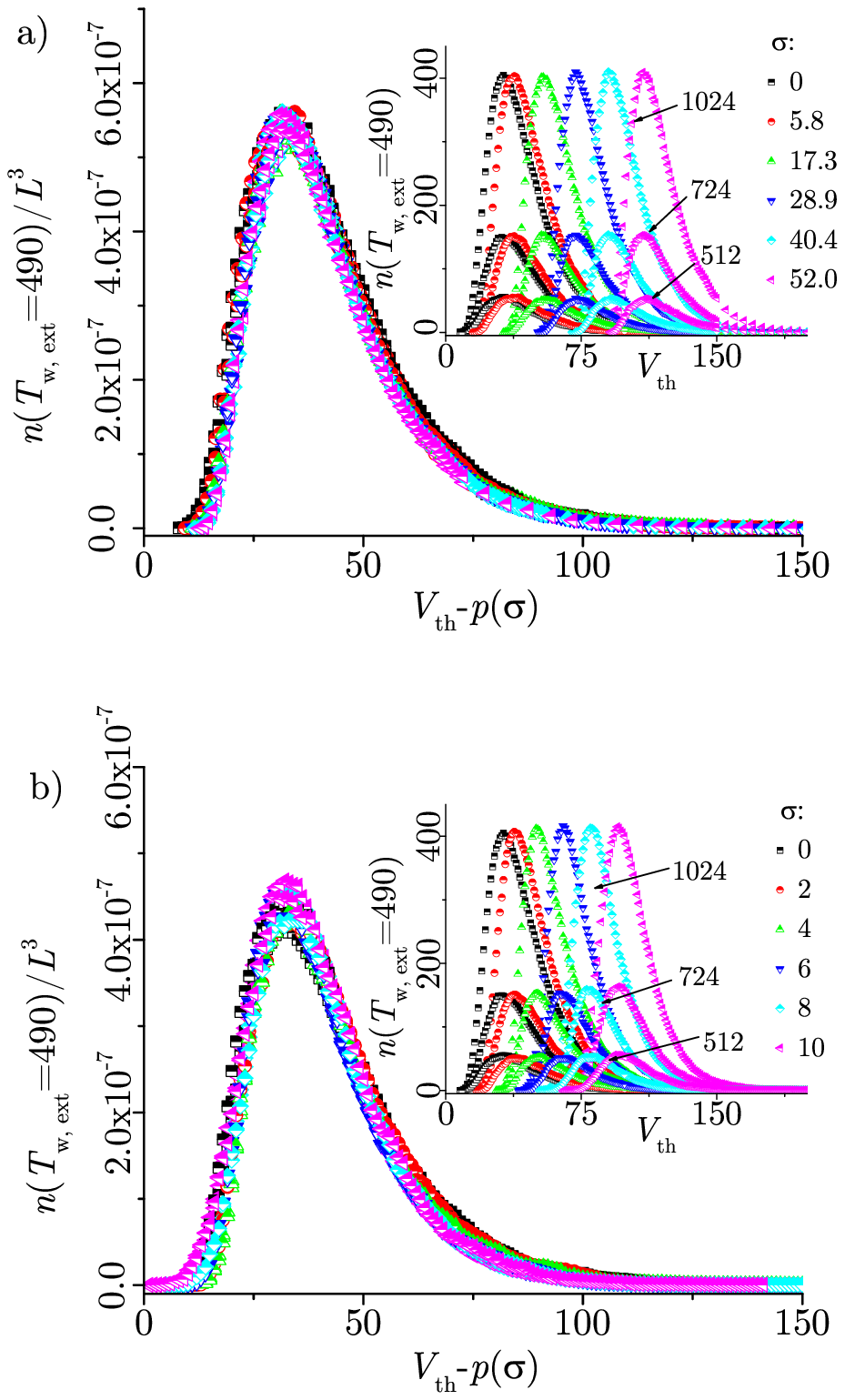}
   \caption{The curves in insets show the distribution of number of occurrences for $T_{\mathrm{w, ext}}=490$ versus $V_{\mathrm{th}}$ for three different systems of linear sizes $L=512$, $L=724$ and $L=1024$ \textcolor{black}{and disorder $R=2.40$}, while in the main panels are presented the collapses of the curves from the insets when (\ref{n(T,V,s,L)}) is applied for the UWN (panel a) and GWN (panel b). Standard deviation ranges from $\sigma=0$ to $\sigma=52$ for UWN, and from $\sigma=0$ to $\sigma=10$ for GWN.}
    \label{Fig8}
\end{figure}

For the fixed $R>R_c$, where $R_c=2.16$ is the critical disorder in the three-dimensional nonequilibrium RFIM \cite{OlgaPRB1999}, the shift parameter is independent on the lattice size $L$ and on the type \textcolor{black}{and length of waiting time, given that waiting time is long enough}. However, \textcolor{black}{$n(T_{\mathrm{w}}; R, V_{\mathrm{th}}, \sigma,L)$} depends on $L$ and should scale with $L^3$, because the number of peaks in the signal scales in that way with the system size. In the main parts of Fig. \ref{Fig8} we present the collapse of the raw $n(\textcolor{black}{T_{\mathrm{w}}=490; R= 2.40, V_{\mathrm{th}},\sigma, L=1024)}$ data from the insets. The collapses are obtained dividing the distribution data by $L^3$ and  translating the threshold values by the shift parameter $p(\sigma)$ corresponding to the UWN data in panel a) and to the GWN data in panel b), respectively. Consequently
\textcolor{black}{
\begin{equation}
  n(T_{\mathrm{w}}; R, V_{\mathrm{th}}, \sigma,L)=L^3\tilde{n}(T_{\mathrm{w}}; R, V_{\mathrm{th}}-p(\sigma)),
  \label{n(T,V,s,L)}
\end{equation} }
where $\tilde{n}(T_{\mathrm{w}}; V_{\mathrm{th}})$ is the scaling function. Complete collapse can be achieved only for sufficiently long waiting times compared to the noise standard deviation. When the examined waiting time is short and $\sigma$ wide, it may happen that there is no such waiting time in the system at all, although it was present for lower values of $\sigma$.

\begin{figure}
    \centering
    \includegraphics[width=0.5\textwidth]{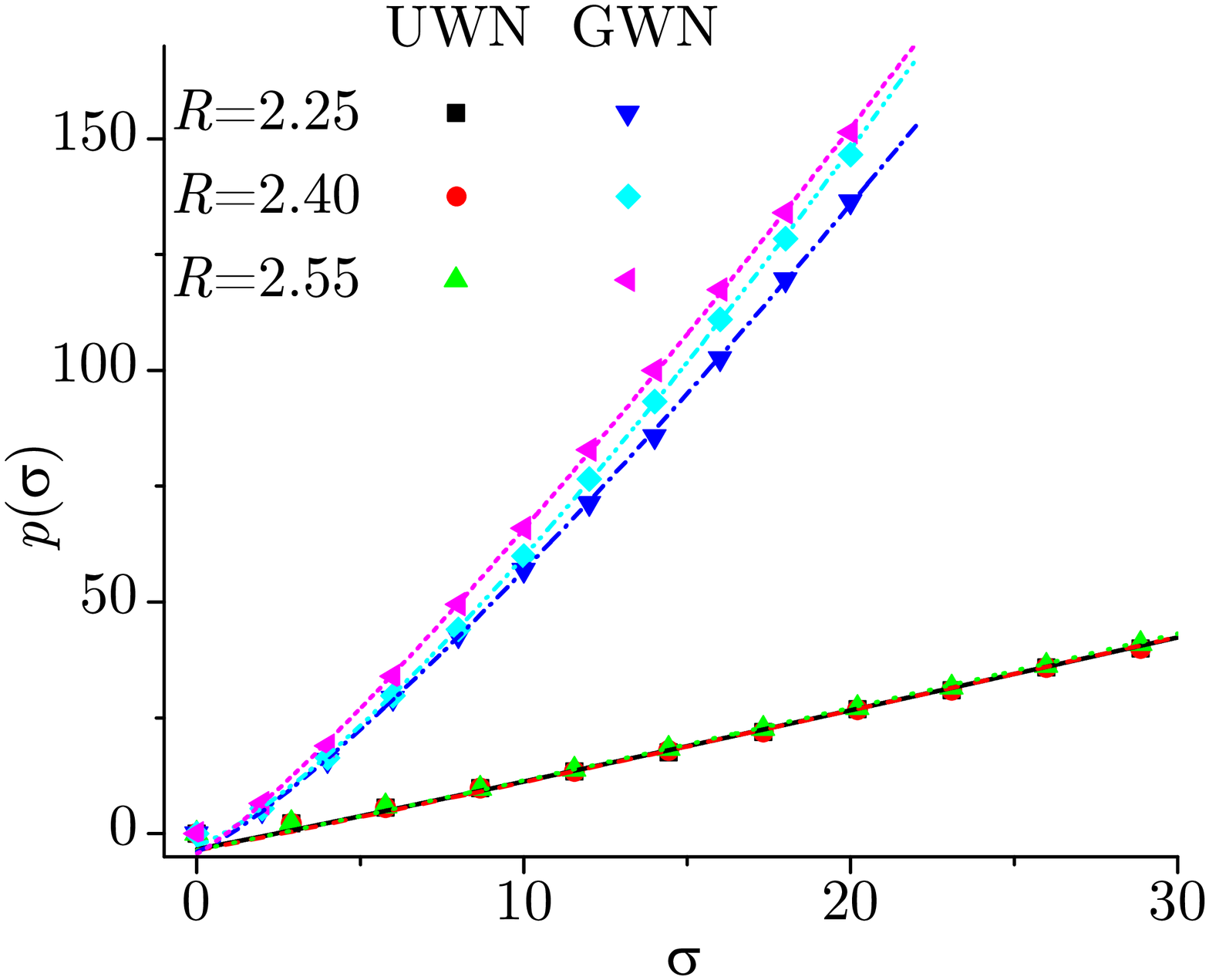}
   \caption{Shift parameter values $p(\sigma)$ obtained from $1024\times1024\times1024$ systems with $R=2.25, 2.40, 2.55$ when UWN and GWN are applied. The data fitted to the function (\ref{shiftfit}) with fitting values presented in tables \ref{Table1} and \ref{Table2}.}
    \label{Fig9}
\end{figure}

It is expected that the shift parameter increases with $\sigma$, since the larger value of threshold is needed to obtain the same value of \textcolor{black}{$n(T_{\mathrm{w}};R,V_{\mathrm{th}},\sigma,L)$} for larger $\sigma$. On the ground of the $p(\sigma)$ values obtained for three different disorders, see Fig. \ref{Fig9}, we assume that the shift parameter obeys a modified power law behavior 

\textcolor{black}{
\begin{equation}
p(\sigma) = a + b \sigma^{c}.
\label{shiftfit}
\end{equation}
}
The fitting curves to this function of several sets of the $p(\sigma)$ data are shown in Fig. \ref{Fig9}. The best fits are obtained for the values of parameters given in Table \ref{Table1} for UWN and in Table \ref{Table2} for GWN. Here one can notice that the parameter values are the same within the error bars for various disorders of UWN, whereas for GWN they significantly depend on disorder.

\begin{table}[!hbt]
    \centering
    \caption{The best fit values of the parameters appearing in (\ref{shiftfit}) in the case of UWN.}
    \begin{tabular}{ c|ccc } 
     \hline
     \hline
        $R$  &  $a$ & $b$ & $c$ \\
     \hline
     \hline
         $2.25$ & $-3.3\pm 0.3$ & $1.33\pm 0.02$ & $1.041\pm 0.002$ \\ 
         $2.40$ & $-3.6\pm 0.3$ & $1.35\pm 0.02$ & $1.038\pm 0.003$ \\ 
         $2.55$ & $-3.6\pm 0.3$ & $1.39\pm 0.02$ & $1.033\pm 0.002$ \\
     \hline
     \hline
    \end{tabular}
    \label{Table1}
\end{table}

\begin{table}[!hbt]
    \centering
    \caption{The same as in Table \ref{Table1}, but for the GWN.}
    \begin{tabular}{ c|ccc } 
     \hline
     \hline
        $R$  &  $a$ & $b$ & $c$ \\
     \hline
     \hline
         $2.25$ & $-3.9\pm 0.9$ & $3.8\pm 0.3$ & $1.20\pm 0.02$ \\ 
         $2.40$ & $-2.6\pm 0.6$ & $3.4\pm 0.2$ & $1.26\pm 0.02$ \\ 
         $2.55$ & $-4.3\pm 0.8$ & $4.8\pm 0.2$ & $1.16\pm 0.02$ \\
     \hline
     \hline
    \end{tabular}
    \label{Table2}
\end{table}

\subsection{Scaling properties}\label{SecScaling}

Both types of waiting time, $T_{\mathrm{w, int}}$ and $T_{\mathrm{w, ext}}$, (jointly denoted by $T_{\mathrm {w}}$) follow the scaling properties of temporal correlations (\ref{scaling}) induced by imposing threshold $V_{\mathrm{th}}$ on the signals obtained from systems with disorders $R$, linear lattice size $L$ and without external noise  \cite{SanjaSciRep}:
\textcolor{black}{
\begin{equation}
\begin{split}
	&V_{\mathrm{th}}^{\frac{\alpha_{\mathrm{int}}\sigma'\nu z}{1-\sigma'\nu z}}D_{T_{\mathrm {w}}}({T_{\mathrm {w}}}; V_{\mathrm{th}}, r, 1/L) = \\ 
	D_{T_{\mathrm {w}}}\Big({T_{\mathrm {w}}}&/V_{\mathrm{th}}^{\frac{\sigma'\nu z}{1-\sigma'\nu z}}; V_{\mathrm{th}}^{\frac{\sigma'^2\nu z}{\sigma'\nu z-1}}r, V_{\mathrm{th}}^{\frac{\sigma'^2\nu^2 z}{\sigma'\nu z-1}}/L\Big).
\end{split}
\label{scaling}
\end{equation}
Here $\alpha$, $\beta$, $\sigma'$, $\nu$, $\delta$, $z$ and $\alpha_{\mathrm{int}}=\alpha+\sigma'\beta\delta/\sigma'\nu z$ are standard RFIM exponents \cite{Sethna2006,OlgaPRB1999,DahmenPRB1999} (note that the standard notation of the, here denoted, exponent $\sigma'$ is $\sigma$, but we choose to denote it here by $\sigma'$ to avoid possible confusion with the noise standard deviation), while $r=(R-R_c)/R$ represents reduced disorder of the system. The exponent $\alpha_{\mathrm{int}}$ is used because the data was gathered from the finite windows of the external magnetic field \cite{SpasojevicPRE2011,SanjaSciRep}.} This means that the distributions $D_{T_{\mathrm {w}}}({T_{\mathrm {w}}}; V_{\mathrm{th}}, r, 1/L)$ of the waiting time, $T_{\mathrm {w}}$ multiplied by \textcolor{black}{$V_{\mathrm{th}}^{\frac{\alpha_{\mathrm{int}}\sigma'\nu z}{1-\sigma'\nu z}}$} and presented versus \textcolor{black}{${{T_{\mathrm {w}}}/V_{\mathrm{th}}^{\frac{\sigma'\nu z}{1-\sigma'\nu z}}}$}, will collapse onto a single curve, if the conditions

\textcolor{black}{
\begin{equation}
V_{\mathrm{th}}^{\frac{\sigma'^2\nu z}{\sigma'\nu z-1}}r=\mathrm{const},\quad V_{\mathrm{th}}^{\frac{\sigma'^2\nu^2 z}{\sigma'\nu z-1}}/L=\mathrm{const},
\label{conditions}
\end{equation} 
}
demanding that systems with the lowest size $L$ have the biggest disorders $R$ and the smallest threshold levels $V_{\mathrm{th}}$, are satisfied.

\begin{figure}
    \centering
    \includegraphics[width=0.5\textwidth]{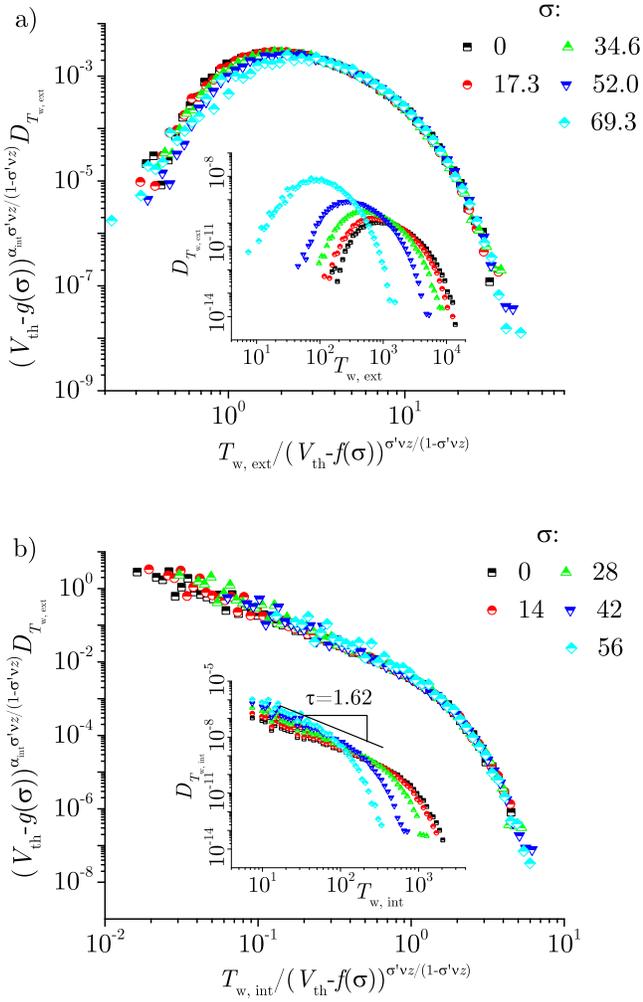}
    \caption{In the main figure of panel a) is presented shifted collapse of the external waiting time distributions obtained from system with $L=2508, R=2.24, V_{\rm {th}}=126$ with added UWN of standard deviation $\sigma$ shown in inset. In panel b) is shown collapse of internal waiting time distributions obtained from the same system with added GWN.}
    \label{Fig10} 
\end{figure}

The addition of external noise alters distributions of waiting time in a way that the increase of the added noise is followed by the rise of short waiting times, while the duration of the longest waiting times decreases, see insets in the Fig. \ref{Fig10}. This means that, upon applying (\ref{scaling}), the waiting time distributions will transform differently so that full collapse will not be achieved. 
In order to obtain the full collapse of distribution of waiting time curves when the external noise is applied
we propose the shift functions $f(\sigma)$ and $g(\sigma)$ that modify scaling relation (\ref{scaling}) into
\textcolor{black}{
\begin{equation}
\begin{split}
	&(V_{\mathrm{th}}-g(\sigma))^{\frac{\alpha_{\mathrm{int}}\sigma'\nu z}{1-\sigma'\nu z}}D_{T_{\mathrm {w}}}({T_{\mathrm {w}}}; V_{\mathrm{th}}, r, 1/L) = \\ 
	D_{T_{\mathrm {w}}}\Big({T_{\mathrm {w}}}&/(V_{\mathrm{th}}-f(\sigma))^{\frac{\sigma'\nu z}{1-\sigma'\nu z}}; V_{\mathrm{th}}^{\frac{\sigma'^2\nu z}{\sigma'\nu z-1}}r, V_{\mathrm{th}}^{\frac{\sigma'^2\nu^2 z}{\sigma'\nu z-1}}/L\Big).
\end{split}
\label{scaling_shift}
\end{equation}}
In this way it is achieved that the transformed distributions overlap with the original distribution obtained without the added noise. In other words, when adequately shifted, noisy distributions behave like the noiseless distributions, see Fig. \ref{Fig10}. There, as in Fig. \ref{Fig13} too, distributions $D_{T_{\mathrm {w}}}({T_{\mathrm {w}}}; V_{\mathrm{th}}, r, 1/L)$ are shortly denoted by $D_{T_{\mathrm{w}}}$.

\begin{figure}
    \centering
    \includegraphics[width=0.5\textwidth]{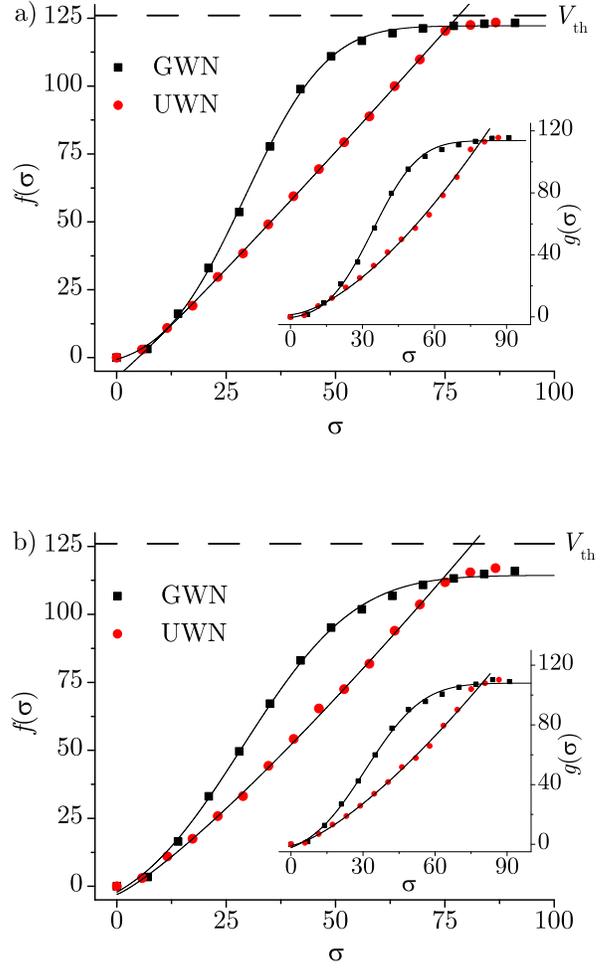}
    \caption{Shift functions $f(\sigma)$ (in main parts) and $g(\sigma)$ (in insets) for the distributions of external waiting time in panel a) and internal waiting time in panel b). The values of shift functions are those used to perform the collapsing shown in main panels of Fig. 10 of the distributions from the insets of \textcolor{black}{that} figure. In the UWN case values are fitted to the power law function $\phi(\sigma) \sim \sigma^s$,
while in the GWN case they are fitted to the error function, $\phi(\sigma)\sim\mathrm{erf}(\sigma)$.}
    \label{Fig11} 
\end{figure}

\begin{figure*}
    \centering
    \includegraphics[width=0.99\textwidth]{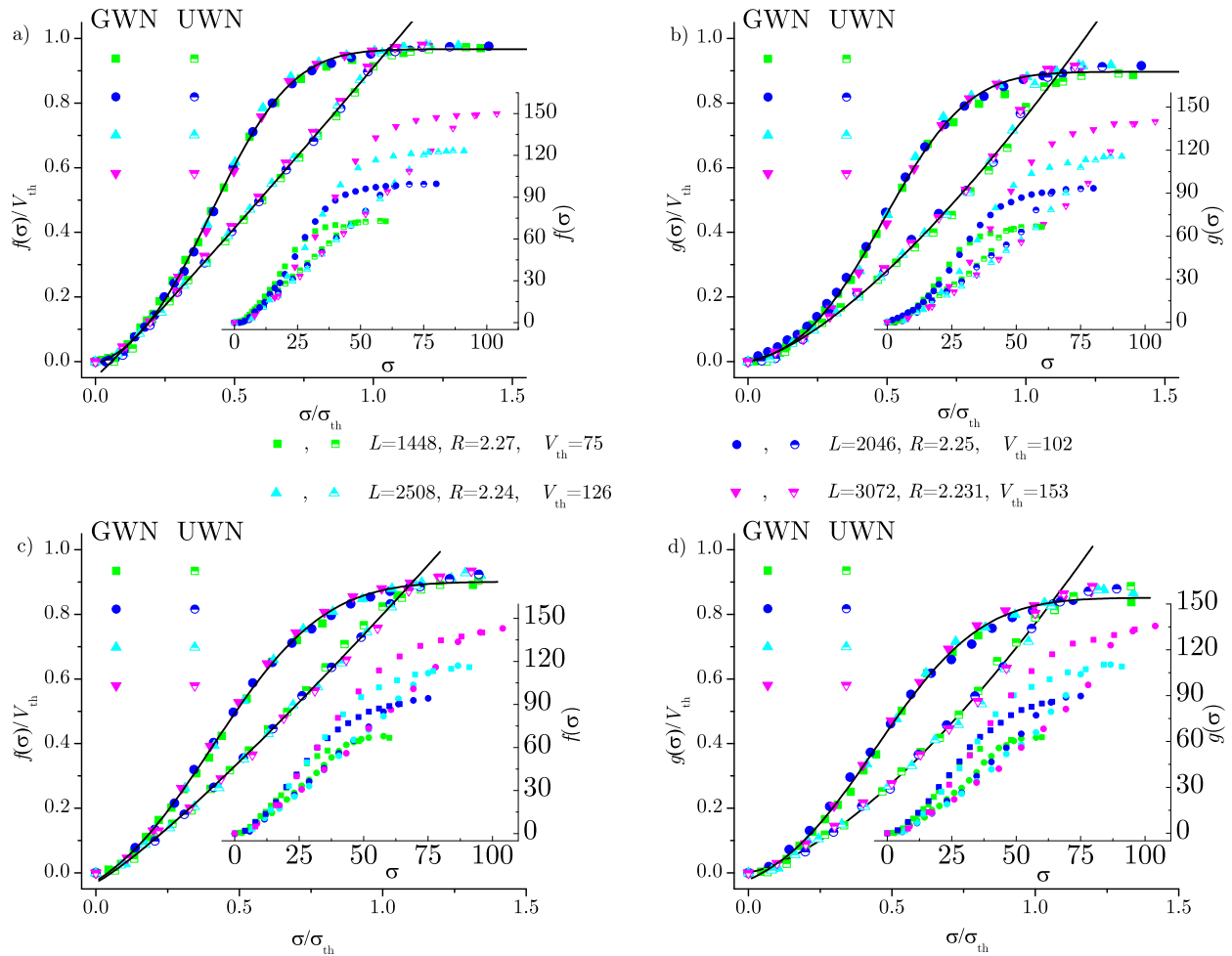}
    \caption{In the insets of panels a) and b) are presented shift functions $f(\sigma)$ and $g(\sigma)$ for the external waiting time distributions obtained for 4 systems such that their dimensions $L$, disorders $R$ and threshold levels $V_{\mathrm{th}}$ satisfy conditions (\ref{conditions}): $L=1448, R=2.27, V_{\mathrm{th}}=75, L=2046, R=2.25, V_{\mathrm{th}}=102, L=2508, R=2.24, V_{\mathrm{th}}=126$, and $L=3072, R=2.231, V_{\mathrm{th}}=153$. On the main panels are shown the collapses of the shift functions divided by $V_{\mathrm{th}}$ as functions of $\sigma/\sigma_{\mathrm{th}}$, fitted to the proposed forms (\ref{shiftfit1}) and (\ref{shiftfit2}). The best fit parameters are given in Tables \ref{Table3} and \ref{Table4}. Panels c) and d): the same as a) and b), but for the internal waiting time.}
    \label{Fig12} 
\end{figure*}

\begin{figure*}
    \centering
    \includegraphics[width=0.99\textwidth]{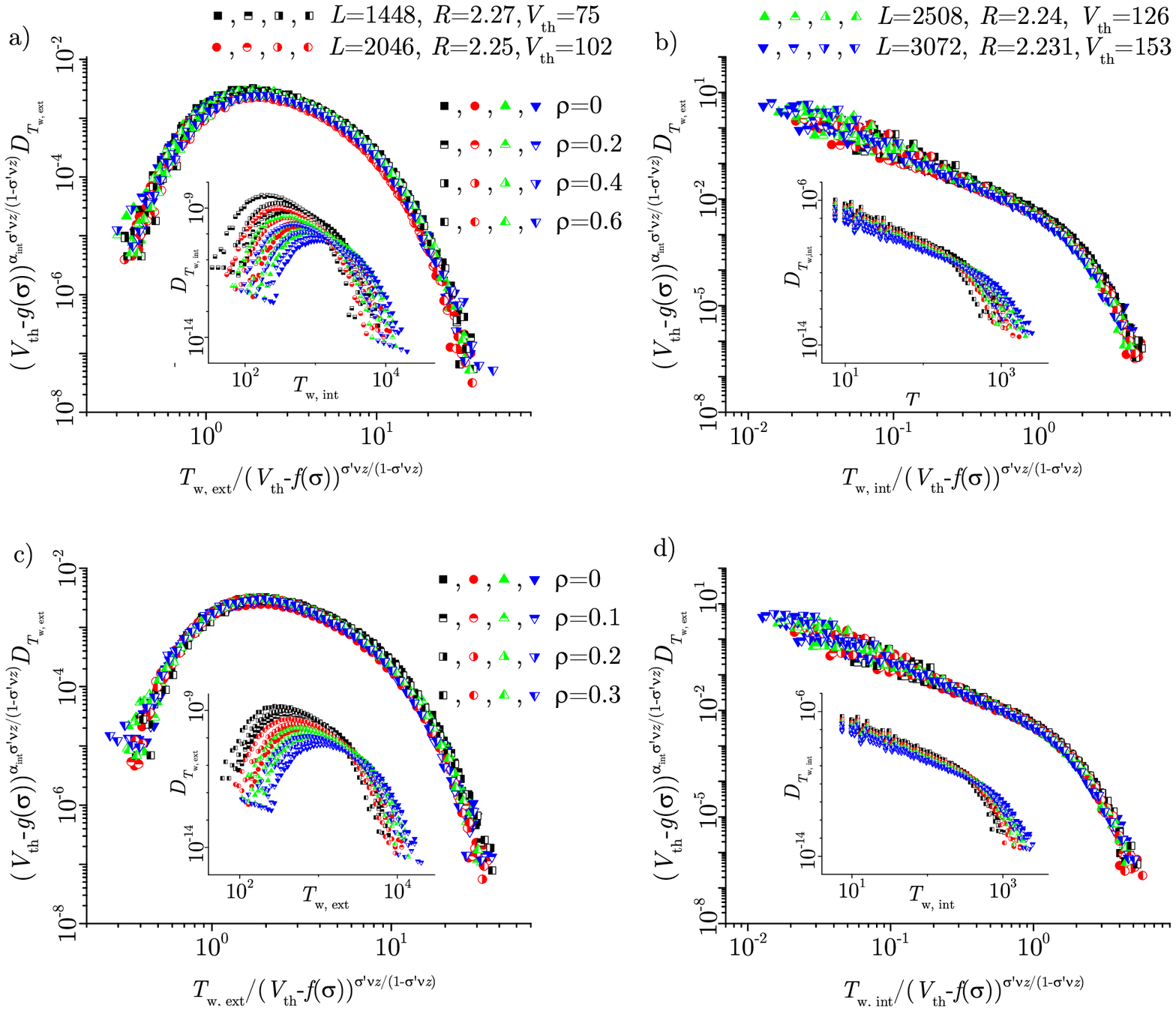}
    \caption{Main figures of panels a) and b) show collapses of the distributions of external (a) and internal (b) waiting times obtained from the same systems as the data shown in Fig. \ref{Fig12} (non-collapsed data is shown in insets) but with such a noise that the ratio between standard deviation of added UWN and imposed threshold $\rho$ ranges from 0 to 0.6. Values of shift functions are calculated using fitting parameters given in Tables \ref{Table3} and \ref{Table4}. The same is presented in panels c) and d), but for GWN and $\rho$ in the range from 0 to 0.3.}  
    \label{Fig13} 
\end{figure*}

An example of shifting functions is presented in Fig. \ref{Fig11}, where one can see that their behavior is not the same for UWN and GWN. \textcolor{black}{The reason for this lies in the difference in the type of noise. Namely, UWN is bounded from the both sides while GWN theoretically can have any value, meaning that in GWN case we need to shift distributions more than in UWN case for the same $\sigma$.} Both $f(\sigma)$ and $g(\sigma)$, when UWN is applied, have a power law behavior $\phi(\sigma)\sim \sigma^s$ ($f(\sigma)$ is almost linear), while in the case when the observed signal contains added GWN their behavior can be described using the error function, $\phi(\sigma)\sim \mathrm{erf}(\sigma)$.
Here it can be seen that after $\sigma$ reaches the value $\sigma_\mathrm{th}\approx V_{\mathrm{th}}/\sqrt{3}$, 
there is almost no difference between $f(\sigma)$ functions in the UWN and GWN case (the same for $g(\sigma)$) 
and that both tend to the value of threshold level $V_{\mathrm{th}}$. 
\textcolor{black}{
In the UWN case, the shift functions reach the threshold value when the noise width $w$ becomes comparable to the value of $V_{\rm th}$ 
(equivalently, when $\sigma\approx V_{\rm th}/\sqrt{3}$), while in the GWN case this happens for a smaller value of $\sigma$.
} 
In the insets of Fig.\ref{Fig12} can be seen that the shift functions $f(\sigma)$ and $g(\sigma)$, 
obtained for 4 different systems follow the same rules: the systems in question satisfy conditions 
(\ref{conditions}): $L=1448, R=2.27, V_{\mathrm{th}}=75; L=2046, R=2.25, V_{\mathrm{th}}=102; 
L=2508, R=2.24, V_{\mathrm{th}}=126$ and $L=3072, R=2.231, V_{\mathrm{th}}=153$. This indicates that 
the functions $f(\sigma)$ and $g(\sigma)$, divided by threshold values $V_{\mathrm{th}}$ and presented as 
functions of $\sigma/\sigma_{\mathrm{th}}$, collapse onto a single curve, meaning that the behavior of the 
shift functions obtained using the set of parameters that satisfy conditions (\ref{conditions}), can be jointly described by
\begin{equation}
    \frac{\phi(\sigma)}{{V_\mathrm{th}}}=p+q\Big(\frac{\sigma}{\sigma_{\mathrm{th}}} \Big)^s
    \label{shiftfit1}
\end{equation}
when UWN is applied, and by
\begin{equation}
    \frac{\phi(\sigma)}{V_{\mathrm{th}}}=i+j\mathrm{erf}\Big(k\frac{\sigma}{\sigma_{\mathrm{th}}}-l\Big)
	\label{shiftfit2}
\end{equation}
in the GWN case. Such collapses are presented in the main panels of Fig. \ref{Fig12} alongside the fitting functions obtained for the values of parameters presented in Tables \ref{Table3} and \ref{Table4}.

\begin{table}[!hbt]
    \centering
    \caption{Values of the best fit parameters for (\ref{shiftfit1}) of the shift functions calculated for internal and external waiting times when the UWN is applied.}
    \begin{tabular}{ c|ccc } 
     \hline
     \hline
       $$ & $p$ & $q$ & $s$ \\
     \hline
     \hline  
     \multicolumn{4}{c}{Internal}\\ 
     \hline   
         $f(\sigma)$ & $-0.028\pm 0.009$ & $0.83\pm 0.01$ & $1.16\pm 0.03$ \\
         $g(\sigma)$ & $ -0.003\pm 0.008$ & $0.77\pm 0.01$ & $1.45\pm 0.05$ \\
     \hline
     \multicolumn{4}{c}{External}\\
     \hline
     	$f(\sigma)$ & $-0.042\pm 0.007$ & $0.96\pm 0.01$ & $1.08\pm 0.02$ \\
        $g(\sigma)$ & $\text{  }0.002\pm 0.008$ & $0.76\pm 0.01$ & $1.44\pm 0.05$\\
     \hline
     \hline
    \end{tabular}
    \label{Table3}
\end{table}

\begin{table}[!hbt]
    \centering
    \caption{Values of the best fit parameters for (\ref{shiftfit2}) of the shift functions calculated for internal and external waiting times when the GWN is applied.}
    \begin{tabular}{ c|cccc } 
     \hline
     \hline
       $$ & $i$ & $j$ & $k$ & $l$ \\
     \hline
     \hline 
     \multicolumn{5}{c}{Internal}\\
     \hline     
         $f(\sigma)$ & $0.391\pm 0.007$ & $0.51\pm 0.01$ & $2.32\pm 0.06$ & $0.94\pm 0.04$ \\
         $g(\sigma)$ & $0.387\pm 0.007$ & $0.46\pm 0.01$ & $2.47\pm 0.08$ & $1.09\pm 0.05$ \\
     \hline
     	\multicolumn{5}{c}{External}\\
     \hline
     	 $f(\sigma)$ & $0.462\pm 0.004$ & $0.505\pm 0.005$ & $3.09\pm 0.05$ & $1.29\pm 0.03$ \\
         $g(\sigma)$ & $0.439\pm 0.004$ & $0.458\pm 0.005$ & $2.96\pm 0.07$ & $1.44\pm 0.04$ \\
     \hline
     \hline
    \end{tabular}
    \label{Table4}
\end{table}
As an illustration, in Fig.\ref{Fig13} are given the collapses obtained for several values of $\rho=\sigma/\sigma_{\mathrm{th}}$, where the shift functions are calculated using the fit parameters from Tables \ref{Table3} and \ref{Table4}.

\begin{figure}
    \centering
    \includegraphics[width=0.5\textwidth]{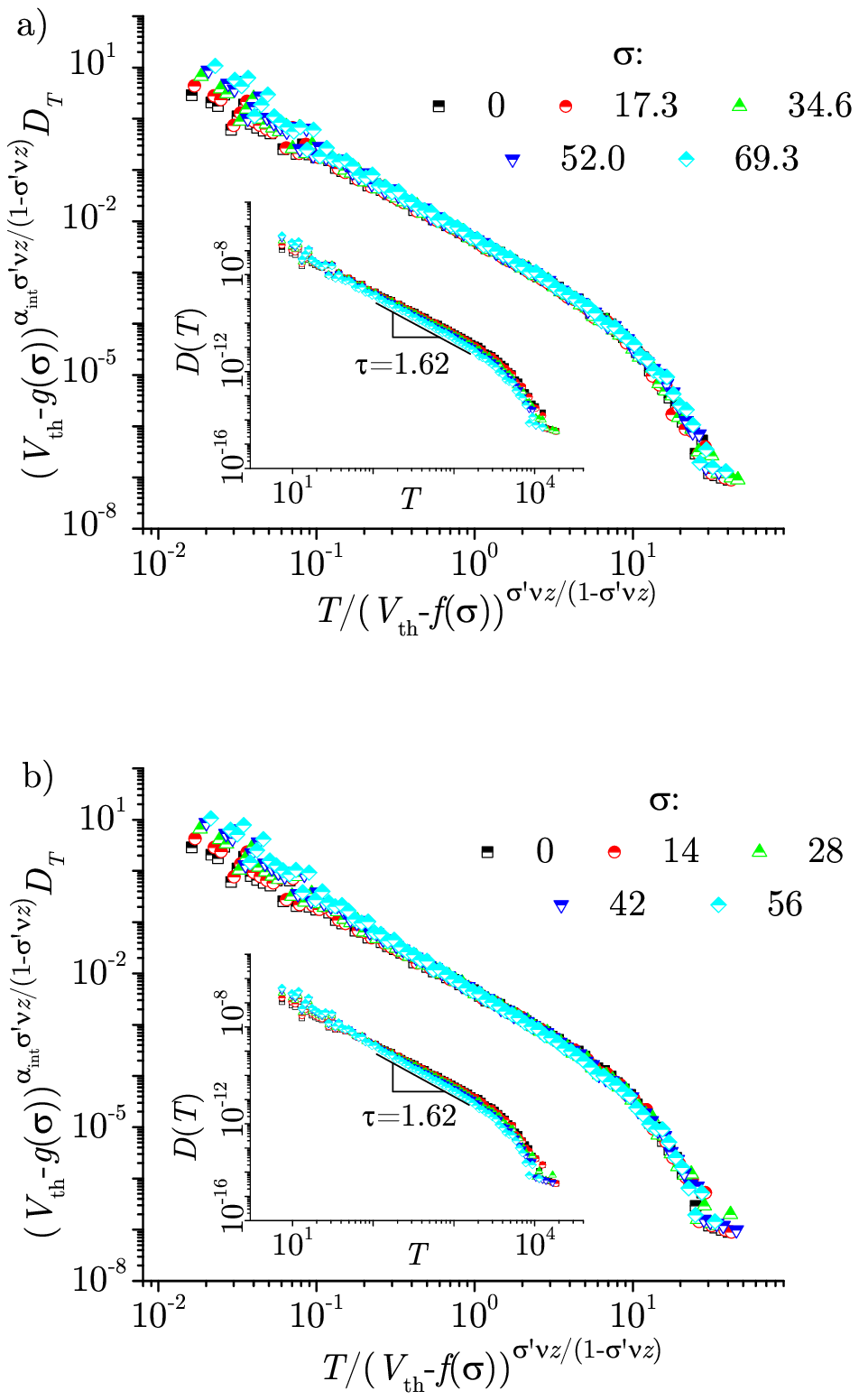}
    \caption{Main figures of panels a) and b) show shifted collapse of the distributions of duration obtained from the same system as the data shown in Fig. \ref{Fig10} when UWN (a) and GWN (b) is added, while non-collapsed 
    \textcolor{black}{distributions, much less affected than the waiting time distributions by the presence of noise from the employed range, are}  shown in insets.}  
    \label{Fig14} 
\end{figure}

\begin{figure}
    \centering
    \includegraphics[width=0.5\textwidth]{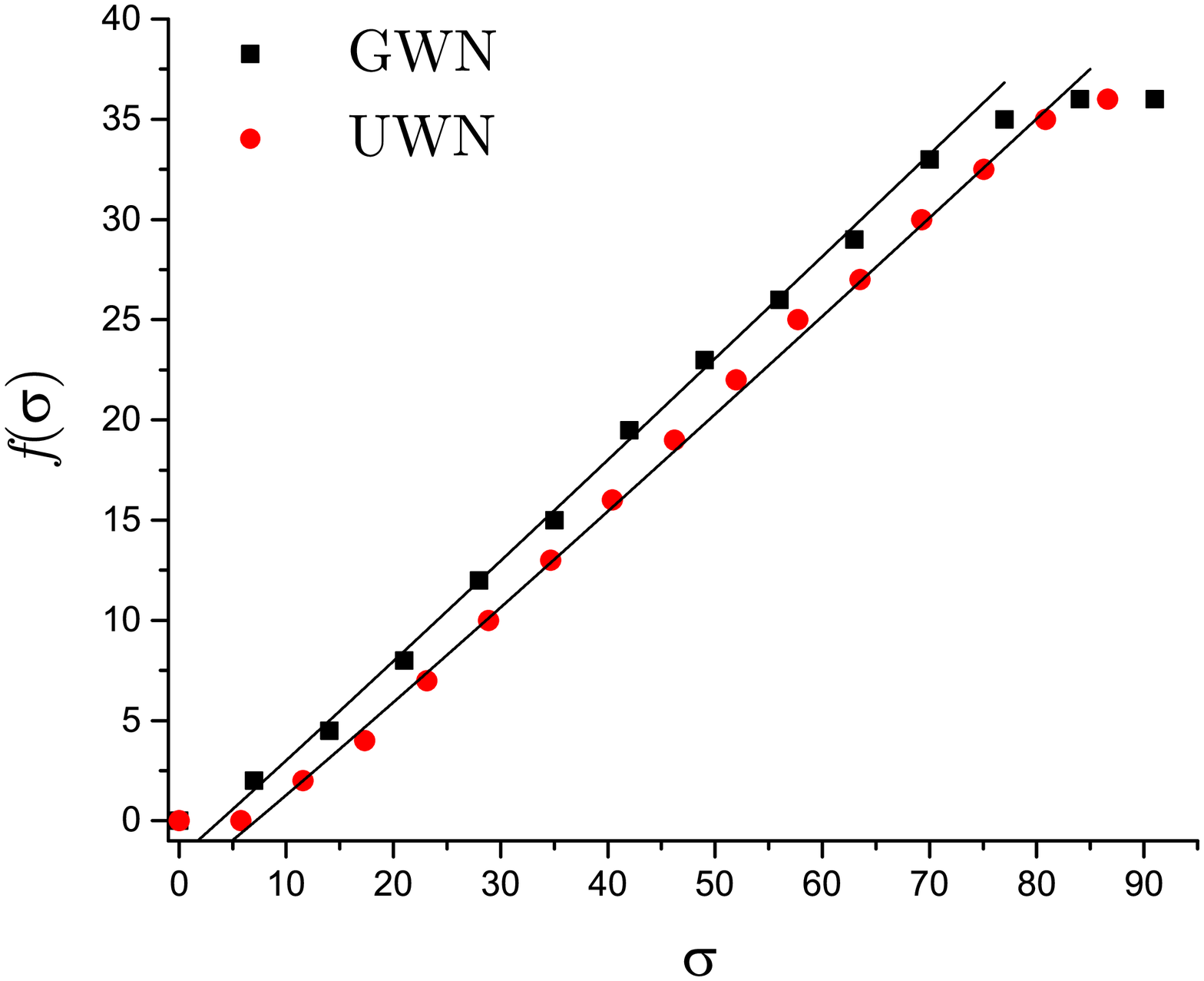}
    \caption{Shift function $f(\sigma)$ for the distributions of duration. The values of shift functions are those used to perform the collapsing shown in main panels of Fig. \ref{Fig14}. In both UWN and GWN cases the values of shift function are fitted to the power law function $\phi(\sigma) \sim \sigma^s$. The values of parameter $s$ are $1.04\pm0.05$ in UWN and $0.96\pm0.06$ GWN case}  
    \label{Fig15} 
\end{figure}

The distribution of avalanche duration $T$ also follows the scaling properties (\ref{scaling}), (with a change in notation $T_{\mathrm{w}}\rightarrow T$) which are affected when the external noise is added, see insets in Fig. \ref{Fig14}. One can see that adding of noise doesn't change significantly these distributions as external noise grows like in the case of waiting times. This holds as long as the threshold level was chosen so that the majority of signal stays above the $V_{\mathrm{th}}$ and the standard deviation of the external noise, comparable to $V_{\mathrm{th}}$, is much smaller than the amplitude of the signal, as is usually accomplished in experiments. In this way the presence of external noise can not significantly decrease the number of events of long duration. Still, the symmetry between distributions of duration and internal waiting time is present since both distributions follow the power-laws, $D_T \sim T^{-\tau_T}$ and $D_{T_{\mathrm{w, int}}}\sim {T_{\mathrm{w, int}}}^{-\tau_{T_{\mathrm{w, int}}}}$ with the same value of exponent $\tau=\tau_T=\tau_{T_{\mathrm{w, int}}}\approx 1.62$ \cite{SanjaPRL}, which is unaffected when the external noise is added. The shifting (\ref{scaling_shift}) can also be applied in the case of duration distributions, see main panels in Fig. \ref{Fig14}, but now the shifting functions behave differently - $g(\sigma)$ is zero, while $f(\sigma)$ is nearly linear for both UWN and GWN, as can be seen in Fig. \ref{Fig15}.

\section{Discussion and Conclusion}\label{conclusion}

In experimental research the impact of external noise and detection threshold is inevitably present. Still, it is often common to pay a minor attention to those impacts in theoretical and numerical analysis of the experimental results. For example, in paper \cite{SanjaSciRep} was shown that the reason for the difference in the experimentally and numerically obtained values of exponent $\gamma$ lies possibly in the effects of nonzero threshold level. Thus, in the present paper we focused on the particular features that are affected by the introduced threshold level and external noise. We observe that the external noise may cause the exponent $\gamma$ to reach its plateau value for lower threshold levels than in the case when there is no external noise. This means that in experiments it would be very difficult to detect the higher values of $\gamma$ before the plateau because the external noise sets a very low threshold level above which the plateau appears. On the other hand, the different noise level impacts the number of occurrences of the given waiting time in the system, see Fig. \ref{Fig6}. Here we see that the interplay between the threshold level and the noise can explain the potential disagreements in the experimental curves of the same type.

In this paper we didn't present the results that correspond to a very large noise, i.e. the noise having magnitude comparable to the average signals. Still, we can see some of the consequences that arise from the large noise, for example in Fig. \ref{Fig5} in the GWN case, where the $\gamma$ values decrease linearly for smaller $\sigma$, but after some value of noise standard deviation the analytical type of decrease changes. This does not happen in the case of UWN, due to the fact that UWN is bounded from both sides, while in GWN case there are no boundaries for the noise values. Thus, although the standard deviations are the same, there is a larger probability of getting greater values for the noise in the GWN case. We expect that the linear drop in the UWN case, presented in Fig. \ref{Fig5}, also proceeds to some other type of decrease, but with much larger values of $\sigma$. Although the effects created by noise would possibly lead to some substantial differences in the presented quantities, it is still not of interest to investigate such noise, because the main motivation for the present research came from the experimental studies, and experimental study of any phenomenon becomes useless if the external (unwanted) noise is that large.

The collapses presented in Section \ref{SecScaling} also show in what way the applied noise disturb the original system's response. We see that the curves of the distributions of waiting time and duration can also collapse onto a single curve when the external noise is applied, but one has to be careful \textcolor{black}{and to modify equation (\ref{scaling}) by adding the shifting functions like in (\ref{scaling_shift})}. Although different types of noise bring quantitatively different effects, these effects are qualitatively the same. This is expected due to the above explained reasons and the manner how the noise affects the signal. 

To conclude, in this paper we examined the joint impact of the external noise and detection threshold level on the response of the externally driven nonequilibrium athermal RFIM. We showed that both noise and threshold level significantly affect the behavior of the signal properties and scaling relations. Thus, the inevitable experimental occasions indeed influence the studied phenomena and should be adequately treated in order to obtain adequate results.

\begin{acknowledgments}
This work was supported by the Serbian Ministry of Education, Science and Technological development.
\end{acknowledgments}


\section*{References}	

\begin{thebibliography}{99}

\bibitem{Earthquakes} J. Davidsen, and M. Baiesi, Phys. Rev. E  \textbf{94}, 022314 (2016).

\bibitem{NeuronalAvalanches2012} N. Friedman, S. Ito, B. A. W. Brinkman, M. Shimono, R. E. L. DeVille, K. A. Dahmen, J. M. Beggs, and  T. C. Butler, Phys. Rev. Lett. \textbf{108}, 208102 (2012).

\bibitem{BrainSignalsPRL2006} C. Bedard, H. Kroger, and A. Destexhe, Phys. Rev. Lett. \textbf{97}, 118102 (2006).

\bibitem{Financial2013} J. P. Bouchaud, J. Stat. Phys. 151-567 (2013).

\bibitem{Zaiser} M. Zaiser, in \textit{Crystal Growth - From
Fundamentals to Technology}, edited by G. Müller, J.-J. Metois and P. Rudolph, p. 215-238 (Elsevier, Amsterdam 2004).

\bibitem{Lase2020} T. Mäkinen, P. Karppinen, M. Ovaska, L. Laurson, and M. J. Alava, Sci. Adv. \textbf{6}, eabc7350 (2020).

\bibitem{Lase2018} H. Salmenjoki, M. J. Alava, and L. Laurson, Nat. Commun. \textbf{9}, 5307 (2018).

\bibitem{Lase2014} P. D. Ispanovity, L. Laurson, M. Zaiser, I. Groma, S. Zapperi, and M. J. Alava, Phys. Rev. Lett. \textbf{112}, 235501 (2014).

\bibitem{SanjaJStat} S. Janicevic, M. Ovaska, M. J. Alava, and L. Laurson, J. Stat. Mech., P07016 (2015).

\bibitem{Budrikis} Z. Budrikis, D. F. Castellanos, S. Sandfeld, M. Zaiser, and S. Zapperi, Nat Commun \textbf{8}, 15928 (2017). 

\bibitem{Sandfeld} S. Sandfeld, Z. Budrikis, S. Zapperi, and D. F. Castellanos, J. Stat. Mech., P02011 (2015).

\bibitem{Santucci} S. Santucci, K.-T. Tallakstad, L. Angheluta, L. Laurson, R. Toussaint, and K. J. Måløy, Phil. Trans. R. Soc. \textbf{A 377}, 20170394 (2018).

\bibitem{SanjaPRL} S. Janicevic, L. Laurson, K. J. Måløy, S. Santucci, and M. J. Alava, Phys. Rev. Lett. \textbf{117}, 230601 (2016).

\bibitem{BelangerNatterman} D. P. Belanger, and T. Nattermann, in \textit{ Spin Glasses and Random Fields}, edited by A. P. Young (World Scientific, Singapore, 1998).

\bibitem{Lieneweg1972} U. Lieneweg, and W. Grosse-Nobis, Inter. J. Magnetism \textbf{3}, 11–16 (1972).

\bibitem{Stanley1996} Dj. Spasojevi\'c, S. Bukvi\'c, S. Milo\v sevi\'c, and H. E. Stanley, Phys. Rev. E \textbf{54}, 2531 (1996).

\bibitem{Durin2000} G. Durin, S. Zapperi, Phys. Rev. Lett.\textbf{84}, 4705 (2000).

\bibitem{Kim2003} D. H. Kim, S. B. Choe, and S. C. Shin, Phys. Rev. Lett. \textbf{90}, 087203 (2003).

\bibitem{Shin2007} S. C. Shin, K. S.Ryu, D. H. Kim, S. B. Choe, and H.Akinaga, J. Magn. Magn. Mater. \textbf{310}, 2599-2603 (2007).

\bibitem{Ryu2007} K. S. Ryu, H. Akinaga, and S-Ch. Shin, Nat. Phys. \textbf{3}, 574 (2007).

\bibitem{Benassi-Zapperi-2011} A. Benassi, and S. Zapperi, Phys. Rev. B  \textbf{84}, 214441 (2011).

\bibitem{Lima2017} G. Z. dos Santos Lima, G. Corso, M. A. Correa, R. L. Sommer, P. Ch. Ivanov, and F. Bohn, Phys. Rev. E \textbf{96}, 022159 (2017).

\bibitem{Bohn2018} F. Bohn, G. Durin, M. A. Correa, N. Ribeiro Machado, R. Domingues Della Pace, C. Chesman, and R. L. Sommer Sci Rep \textbf{8}, 11294 (2018).

\bibitem{Cizeau1998} P. Cizeau, S. Zapperi, G. Durin, and H. E. Stanley, Phys. Rev. Lett. \textbf{79}, 4669 (1997).

\bibitem{Zapperi1998} S. Zapperi, P. Cizeau, G. Durin, and H. E. Stanley, Phys. Rev. B \textbf{58}, 6353 (1998).

\bibitem{Sethna2006} J. P. Sethna, K. A. Dahmen, O. Perkovi\'c, in \textit{The Science of Hysteresis}, edited by G. Bertotti and I. Mayergoyz (Academic, Amsterdam, 2006).


\bibitem{SethnaJMMM2001} K. A. Dahmen, J. P. Sethna, M. C. Kuntz, and O. Perkovi\'c, J. Magn. Magn. Mater. \textbf{226}, 1287 (2001).

\bibitem{Bertotti1990} G. Bertotti, and M. Pasquale, J. App. Phys. \textbf{67}, 5255 (1990).

\bibitem{Franz2011} S. Franz, G. Parisi, F. Ricci-Tersenghi, and T. Rizzo, Eur. Phys. J. E \textbf{34}, 102 (2011).

\bibitem{Vives1994} E. Vives, and A. Planes, Phys. Rev. B \textbf{50}, 3839 (1994).

\bibitem{VivesJMMM2000} E. Vives, and A. Planes, J. of Magn. and Magn. Mat. \textbf{221}, 164-171 (2000).

\bibitem{ABBM} B. Alessandro, C. Beatrice, G. Bertotti, and A. Montorsi, J. Appl. Phys. \textbf{68}, 2901 (1990); \textbf{68}, 2908 (1990).

\bibitem{DW} S. Zapperi, P. Cizeau, G. Durin, and H. E. Stanley, Phys. Rev. B \textbf{58}, 6353 (1998).

\bibitem{LassePRB2014} L. Laurson, G. Durin, S. Zapperi, Phys. Rev. B \textbf{89}, 104402 (2014).

\bibitem{Schulz1988} U. Schulz, J. Villain, E. Brézin, and H. Orland, J. Stat. Phys. \textbf{51}, 1 (1988).

\bibitem{TadicPRL1996} B. Tadi\'c, Phys. Rev. Lett. \textbf{77}, 3843 (1996).	

\bibitem{BalogPRB2018} I. Balog, G. Tarjus, and M. Tissier, Phys. Rev. B \textbf{97}, 094204 (2018).

\bibitem{SethnaPRL93}   J. P. Sethna, K. Dahmen, S. Kartha, J. A. Krumhansl, B. W. Roberts, and J. D. Shore,  Phys. Rev. Lett. \textbf{70}, 3347 (1993)

\bibitem{Vives2001} E. Vives, and A. Planes, Phys. Rev. B \textbf{63}, 134431 (2001).

\bibitem{FytasPRL2013} N. G. Fytas, and V. Martin-Mayor, Phys. Rev. Lett. \textbf{110}, 227201 (2013).

\bibitem{JSTAT2021}  S. Jani\'{c}evi\'{c}, D. Kne\v{z}evi\'{c}, S. Mijatovi\'{c}, and D. Spasojevi\'{c},  J. Stat. Mech., 013202 (2021).

\bibitem{YoungPerturbation} A. P. Young, J. Phys. A: Math. Gen. \textbf{10}, L257 (1977).

\bibitem{ParisiPerturbation} G. Parisi, and N. Sourlas, Phys. Rev. Lett. \textbf{43}, 744 (1979).

\bibitem{BricmontPerturbation} J. Bricmont, and A. Kupiainen, Phys. Rev. Lett. \textbf{59}, 1829 (1987).

\bibitem{ParisiNonPerturbation} G. Parisi, and N. Sourlas, Phys. Rev. Lett. \textbf{89}, 257204 (2002).

\bibitem{TissierNonPerturbation} M. Tissier, and G. Tarjus, Phys. Rev. Lett. \textbf{107}, 041601 (2011).

\bibitem{FytasPRL2016} N. G. Fytas, V. Martin-Mayor, M. Picco, and N. Sourlas, Phys. Rev. Lett. \textbf{116}, 227201 (2016).

\bibitem{FytasPRE2017} N. G. Fytas, V. Martin-Mayor, M. Picco, and N. Sourlas, Phys. Rev. E \textbf{95}, 042117 (2017).

\bibitem{FytasPRL2019} N. G. Fytas, V. Martin-Mayor, G. Parisi, M. Picco, and N. Sourlas, Phys. Rev. Lett. \textbf{122}, 240603 (2019).

\bibitem{OlgaPRB1999} O. Perkovi\'c, K. A. Dahmen, and J. P. Sethna, Phys. Rev. B \textbf{59}, 6106 (1999).

\bibitem{DahmenPRB1999} K. A. Dahmen, and J. P. Sethna, Phys. Rev. B \textbf{53}, 14872 (1996). 

\bibitem{RechePRB2003} F. J. Perez-Reche and E. Vives, Phys. Rev. B \textbf{67}, 134421 (2003).

\bibitem{RechePRB2004}  F. J. Perez-Reche and E. Vives, Phys. Rev. B \textbf{70}, 214422 (2004).

\bibitem{SpasojevicEPL2006}  Dj. Spasojevi\'c, S. Jani\'cevi\'c, and M. Kne\v{z}evi\'c, Europhys. Lett.  \textbf{76}, 912 (2006).

\bibitem{SpasojevicPRL2011} Dj. Spasojevi\'c, S. Jani\'cevi\'c, and M. Kne\v zevi\'c, Phys. Rev. Lett. \textbf{106}, 175701 (2011).

\bibitem{SpasojevicPRE2011} Dj. Spasojevi\'c, S. Jani\'cevi\'c, and M. Kne\v zevi\'c, Phys. Rev. E \textbf{84}, 051119 (2011).

\bibitem{SpasojevicPRE2014} Dj. Spasojevi\'c, S. Jani\'cevi\'c, and M. Kne\v zevi\'c, Phys. Rev. E. \textbf{89}, 012118 (2014).

\bibitem{NavasVives} V. Navas-Portella, and E. Vives, Phys. Rev. E \textbf{93}, 022129 (2016).

\bibitem{3D-2D2018} Dj. Spasojevi\'c, S. Mijatovi\'c, V. Navas-Portella, and E. Vives,  Phys. Rev. E \textbf{97}, 012109 (2018).

\bibitem{BosaSciRep2019} B. Tadi\'c, S. Mijatovi\'c, S. Jani\'cevi\'c, Dj. Spasojevi\'c, G. J. Rodgers, Sci. Rep. \textbf{9}, 6340 (2019).

\bibitem{3D-2DHceff} S. Mijatovi\'c, D.Jovkovi\'c, S. Jani\'cevi\'c, and Dj. Spasojevi\'c,  Phys. Rev. E \textbf{100}, 032113 (2019).

\bibitem{SvetaPRE2020} S. Mijatovi\'c, M. Brankovi\'c, S. Graovac, and Dj. Spasojevi\'c,  Phys. Rev. E \textbf{102}, 022124 (2020).

\bibitem{Shukla2013PRE} D. Thongjaomayum and P. Shukla, Phys. Rev. E \textbf{88}, 042138 (2013)

\bibitem{Shukla2015PRE} L. Kurbah, D. Thongjaomayum and P. Shukla, Phys. Rev. E \textbf{91}, 012131 (2015).

\bibitem{Shukla2016} P. Shukla, and D. Thongjaomayum, J. Phys. A: Math. Theor. \textbf{49}, 235001 (2016).

\bibitem{Triangle} S. Jani\'cevi\'c, S. Mijatovi\'c, and Dj. Spasojevi\'c, Phys. Rev. E \textbf{95}, 042131 (2017).

\bibitem{Shukla2019} D. Thongjaomayum and P. Shukla, Phys. Rev. E \textbf{99}, 062136 (2019).

\bibitem{BohnExp} I. P. de Sousa, G. Z. dos Santos Lima, M. A. Correa, R. L. Sommer, G. Corso, F. Bohn, Sci. Rep. \textbf{10}, 9692 (2020)

\bibitem{SanjaSciRep} S. Jani\' cevi\' c, D. Jovkovi\' c, L. Laurson, Dj. Spasojevi\' c, Sci. Rep. \textbf{8}, 2571 (2018).

\textcolor{black}{
\bibitem{bares2019seismiclike} J. Bar{\'e}s, D. Bonamy, and A. Rosso, Phys. Rev. E \textbf{100}, 023001 (2019)
\bibitem{post2021interevent} R. A. J. Post, M. A. J. Michels, J.-P. Ampuero, T. Candela, P. A. Fokker, J.-D. van Wees, R. W. van der Hofstad, and E. R. van den Heuvel, Edwin R, Sci. Rep. \textbf{11}, 1-10 (2021)
\bibitem{radiguet2016triggering} M. Radiguet, H. Perfettini, N. Cotte, A. Gualandi, B. Valette, V. Kostoglodov, T. Lhomme, A. Walpersdorf, E. C. Cano, M. and Campillo, Nat. Geosci. \textbf{9}, 829 (2016)
\bibitem{font2015perils} F. Font-Clos, G. Pruessner, N. Moloney, and A. Deluca, New J. Phys. \textbf{17}, 043066 (2015) 
}
\textcolor{black}{
\bibitem{Weissman} M. B. Weissman, Rev. Mod. Phys. \textbf{60}, 537 (1988)
\bibitem{Kuntz2000}  M. C. Kuntz and J. P. Sethna, Phys. Rev. B \textbf{62}, 11699 (2000)
}
\bibitem{OlgaCondMat96} O. Perkovi\'c, K. A. Dahmen, and J. P. Sethna, arXiv:cond-mat/9609072 v1 (6 Sep 1996)

\bibitem{Kuntz1999} M. C. Kuntz, O. Perkovi\'c, K. A. Dahmen, B. W. Roberts, J. P. Sethna, Computing in Science and Engineering \textbf{1}, 4 (1999).

\bibitem{AdiabaticReg} In adiabatic regime the interval of time between the end of previous and the beginning of next avalanche is technically infinite because this regime is the limit of the constant driving rate regime when the driving rate $\omega$ tends to zero. For this reason the external waiting in adiabatic regime doesn’t include the time when the system is inactive between two neighboring avalanches surpassing the chosen threshold level \cite{SanjaSciRep}. Nevertheless, one can mimic the finite driving regime at very slow driving by specifying some very small value for $\omega$,  and add to the external waiting time the duration of the system inactivity intervals recalculated from the corresponding change of the external magnetic field.

\bibitem{LasseJStat} L. Laurson, X. Illa, and M. J. Alava, J. Stat. Mech., P01019 (2009).

\end{thebibliography}
\end{document}